\pdfoutput=1
\documentclass[aps,physrev,twocolumn,superscriptaddress,longbibliography]{revtex4-1}
\usepackage[english]{babel}
\usepackage{amsmath, amssymb, mathtools, amsthm, bbm}
\usepackage{graphicx}
\usepackage[babel]{csquotes}
\usepackage{natbib}
\usepackage{url}
\usepackage[colorlinks]{hyperref}
\newcommand{\kB}{k_{\rm B}}

\theoremstyle{plain}
\newtheorem*{FDT}{Generalized FDT}

\begin{document}

\title{A framework of nonequilibrium statistical mechanics. I. Role and types of fluctuations}

\author{Hans Christian \"{O}ttinger}
\affiliation{ETH Z\"{u}rich, Department of Materials, Polymer Physics, CH-8093 Z\"urich, Switzerland}
\author{Mark A. Peletier}
\affiliation{Technische Universiteit Eindhoven, Centre for Analysis, Scientific Computing and Applications, and Institute for Complex Molecular Systems (ICMS), 5600 MB Eindhoven, The Netherlands}
\author{Alberto Montefusco}
\affiliation{ETH Z\"{u}rich, Department of Materials, Polymer Physics, CH-8093 Z\"urich, Switzerland}

\date{\today}

\begin{abstract}
Understanding the fluctuations by which phenomenological evolution equations with thermodynamic structure can be enhanced is the key to a general framework of nonequilibrium statistical mechanics. These fluctuations provide an idealized representation of microscopic details. We consider fluctuation-enhanced equations associated with Markov processes and elaborate the general recipes for evaluating dynamic material properties, which characterize force-flux constitutive laws, by statistical mechanics. Markov processes with continuous trajectories are conveniently characterized by stochastic differential equations and lead to Green-Kubo-type formulas for dynamic material properties. Markov processes with discontinuous jumps include transitions over energy barriers with the rates calculated by Kramers. We describe a unified approach to Markovian fluctuations and demonstrate how the appropriate type of fluctuations (continuous versus discontinuous) is reflected in the mathematical structure of the phenomenological equations.
\end{abstract}

\maketitle

\section{Introduction}
Phenomenological evolution equations with a thermodynamic structure can be enhanced by adding fluctuations. These fluctuations represent the effect of more microscopic, fast degrees of freedom that have been neglected in the phenomenological equations. Whether the fluctuations cause only small corrections or have far-reaching consequences depends on the particular problem of interest.

A simple but important example for the benefits of fluctuation enhancement is provided by the motion of small particles suspended in fluids. If the suspended particles are sufficiently small, such as pollen particles in water, one observes a wild random motion of the particles, which is the famous Brownian motion resulting from incessant collisions with small molecules in the fluid. If the fluid is described by the phenomenological hydrodynamic equations, which specify the evolution of density, velocity and temperature fields, Brownian motion cannot be explained. By adding fluctuations to phenomenological hydrodynamics, one incorporates important features of the molecular nature of the fluid and Brownian motion becomes accessible. An important experimental technique known as microbead rheology (see, for example, Chapt.~25 of the textbook \cite{hcomctp} and references therein) is actually based on the observation of the random motion of small particles in complex fluids, from which the viscoelastic fluid properties can be inferred.

An even more striking example for the importance of fluctuations is dynamic light scattering by a low-molecular-weight liquid, say water (see, for example, the monograph \cite{BernePecora} or Chapt.~26 of the textbook \cite{hcomctp}). This experimental technique is sensitive to time-dependent correlations in mass density on the length scale of the wavelength of light. Typical length scales are below $10^{-7}\,{\rm m}$, typical time scales can go down to $10^{-10}\,{\rm s}$, where these two scales are related by the speed of sound. Ultimately, it is the polarizability of molecules that matters for the scattering of electromagnetic waves. Information on such short length scales is not available in the phenomenological hydrodynamic equations. By adding fluctuations to hydrodynamics, we obtain precisely the small-scale information about density correlations that is required to describe dynamic light scattering \cite{OrtizSengers}. This example demonstrates the usefulness of fluctuating hydrodynamics impressively.

The standard reference for fluctuating hydrodynamics is the textbook by Landau and Lifshitz \cite{LandauLifshitz9}. To respect conservation laws, Gaussian fluctuations are introduced for the fluxes of conserved quantities rather than the conserved quantities themselves. At any time, the local-equilibrium fluctuations are characterized by Einstein's fluctuation theory obtained from equilibrium statistical mechanics (see, for example, Sec.~10.B of \cite{Reichl}). According to Onsager's regression hypothesis \cite{Onsager31a,Onsager31b}, the decay of fluctuations is governed by the phenomenological hydrodynamic equations.

In the context of numerical solutions of hydrodynamic equations, the relevance of fluctuations in a large number of situations has been compiled and discussed by Donev and coworkers. Diffusive transport is strongly enhanced by thermal velocity fluctuations \cite{Donevetal11}. Fluctuations also have a strong impact on the spinodal decomposition of multi-component systems \cite{Chaudhrietal14}. Moreover, fluctuations in reactive systems have a strong effect on giant long-range correlated concentration fluctuations, accelerate pattern formation in spatially homogeneous systems and lead to a qualitatively different disordered pattern behind a traveling wave \cite{Bhattacharjeeetal15,Kimetal17}.

Can hydrodynamic fluctuations in multi-component systems be described in the same way as those in one-component systems? If we consider only the resulting diffusion effects, the answer is yes. If we moreover allow for chemical reactions between the different components, the answer is no. In this paper, building upon the works \cite{Mielkeetal14,Mielkeetal16}, we first show how different types of noise correspond to different thermodynamic structures (force-flux constitutive laws) of the phenomenological equations for isolated systems. Then, as our central contribution, we present the consequences of this correspondence for nonequilibrium statistical mechanics, by answering the following two questions.
(i) Given a microscopic model and a meaningful coarse-graining map, how can one find the thermodynamic structure of the macroscopic, phenomenological equation?
(ii) In the opposite direction, given a phenomenological equation and its thermodynamic structure, how can we add fluctuations?
For these procedures we use the terms ``coarse-graining'' and ``noise enhancement''.

This paper pursues the following line of thought. We start from a class of phenomenological evolution equations for isolated systems with a particular thermodynamic structure (Sec.~\ref{secthermostructure}). To enhance these phenomenological equations, we do not restrict ourselves to continuous Gaussian noise but rather allow for general Markov processes, including jump processes. We then try to recognize the proper types of noise to be used for enhancing phenomenological equations in the thermodynamic structure of those equations. This attempt leads us to the very general fluctuation-dissipation theorem that is at the heart of this paper (Sec.~\ref{secFDT}). Finally, we analyze the noise resulting from more microscopic descriptions to extract the detailed form of more macroscopic, phenomenological evolution equations, which is a key task of nonequilibrium statistical mechanics (Sec.~\ref{secstatmech}).

For the theory of systems in thermodynamic equilibrium, which do not macroscopically evolve in time, we know that the general recipes for calculating thermodynamic potentials by statistical mechanics have not yet been \emph{derived} in any rigorous manner: for instance, there is no definite proof that the thermodynamic entropy, which is a macroscopically measurable quantity \cite{BZ11}, is the same entropy computed by statistical mechanics (for an attempt in this direction, see \cite{WKFR16}). Nevertheless, equilibrium statistical mechanics has been \emph{formulated} in such an elegant and convincing way by Gibbs, whose powerful formulation is based on ideas of Boltzmann and Maxwell, that it is now generally accepted and used with greatest confidence. For nonequilibrium systems, we certainly cannot expect anything better than compelling heuristic arguments and an elegant mathematical formulation of the recipes that can be used to extract the thermodynamic structure of phenomenological evolution equations from microscopic dynamics. In particular, the announced generalized fluctuation-dissipation theorem does not have the status of a mathematical theorem, but rather of a plausible postulate or axiom.

The main goal of all our efforts is providing the tools for condensing experimental observations, partial understanding, statistical arguments, and intuition into a consistent multiscale description for a given problem of interest. For example, available background knowledge should be used to guide the focus of statistical mechanics and to make its procedures more efficient, and intuition about molecular processes should guide appropriate phenomenological modeling even if systematic statistical mechanics is prohibitively expensive. We strive for a multiscale approach with reassuring mutual consistency between different levels of description.

\section{A framework for nonequilibrium thermodynamics}\label{secthermostructure}
To carry out the program sketched in the introduction, it is essential to build on a sound framework for nonequilibrium thermodynamics. On the one hand, if we wish to enhance a phenomenological equation by adding fluctuations, we must be sure that the phenomenological equation is thermodynamically consistent so that a recipe for introducing thermal fluctuations can make physical sense. On the other hand, a major goal of statistical mechanics is to evaluate expressions for the building blocks of a thermodynamic description from a more microscopic description. At equilibrium, we need to find a single thermodynamic potential that characterizes all the thermodynamic equations of state of the system. Nonequilibrium thermodynamics requires more.

Our discussion is based on the GENERIC (\underbar{g}eneral \underbar{e}quation for the \underbar{n}on\underbar{e}quilibrium \underbar{r}eversible-\underbar{i}rreversible \underbar{c}oupling) formulation of time-evolution equations for nonequilibrium isolated systems \cite{hco99,hco100,hcobet,PavelkaKlikaGrmela}, where two different formulations of the fundamental time evolution have already been offered in Eqs.~(4) and (13) of the original paper \cite{hco99}. Both formulations provide autonomous evolution equations for a list of independent variables $x$ that describe an isolated nonequilibrium system of interest, and both formulations rely on an additive superposition of reversible and irreversible contributions to dynamics. The fact that the GENERIC framework deals with isolated systems represents no essential limitation, since the equations for local field theories, the most frequent models for real system, are independent of boundary conditions; in other cases, it may be necessary to include the environment in the description. Attempts to generalize the GENERIC framework to open systems have been made in \cite{hco156,hco162,hco171}.

For simplicity, we here assume that the list of variables $x$ is finite, $x \in \mathbb{R}^d$. The first formulation of GENERIC in \cite{hco99} is given by the equation
\begin{equation} \label{LMformulation}
  \frac{dx}{dt} = L(x) \frac{\partial E(x)}{\partial x} + M(x) \frac{\partial S(x)}{\partial x} \, .
\end{equation}
The quantities $E$ and $S$ are the total energy and entropy as functions of the system variables $x$, and $L$ and $M$ are certain linear operators, or matrices, which are also allowed to depend on $x$. The two contributions to the time evolution of $x$ generated by the energy $E$ and the entropy $S$ in Eq.~(\ref{LMformulation}) are the reversible and irreversible contributions, respectively.  We refer to $L$ and $M$ as the Poisson and friction matrices.

The GENERIC framework postulates a number of properties for the four building blocks $E$, $S$, $L$, $M$, most of which are related to physically sound balance laws for energy and entropy (which, in turn, are associated with the fundamental laws of thermodynamics). The practical advantage of restricting ourselves to isolated systems is that conservation laws can be formulated more easily. The conservation of energy by reversible dynamics is guaranteed by requiring the Poisson matrix $L$ to be antisymmetric. The conservation of energy by irreversible dynamics is assumed for all choices of the generator $S$; this assumption means that $\partial E/\partial x$ is a left eigenvector of the matrix $M$ with eigenvalue zero, thus implying degeneracy of the friction matrix $M$. The conservation of entropy by reversible dynamics, which actually is a hallmark of reversibility, is assumed for all choices of the generator $E$; this assumption means that $\partial S/\partial x$ is an eigenvector of the matrix $L$ with eigenvalue zero, thus implying degeneracy of the Poisson matrix $L$, too. Finally, the friction matrix $M$ is assumed to be positive-semidefinite. This assumption is a very strong formulation of the second law of thermodynamics because it implies that the irreversible contribution to the rate of change of $S$, that is $dS/dt$, is nonnegative for any choice of $S$, not just for the physical entropy.

There are further properties of $L$ and $M$ that are not related to the balance laws for energy and entropy. While the Poisson matrix $L$ has to be antisymmetric, the symmetry properties of the friction matrix $M$ are less obvious. Often $M$ is assumed to be symmetric, but requiring Onsager-Casimir symmetry
is more appropriate (see Sections 3.2.1 and 7.2.4 of \cite{hcobet} as well as \cite{hco218}). A highly restrictive condition on the matrix $L$ is given by the Jacobi identity for the Poisson bracket associated with $L$, which is defined by
$\{A,B\}= (\partial A/\partial x) \cdot L \, (\partial B/\partial x)$. This property expresses the time-structure invariance of reversible dynamics \cite{hco102}. The version (\ref{LMformulation}) of GENERIC is also known as a metriplectic structure \cite{Morrison86}. Note, however, that the ``metric'' $M(x)$ is degenerate and may be non-symmetric (for example, in modeling turbulence \cite{hco218} or slip \cite{hco143}).

The second formulation of GENERIC in \cite{hco99} is given by the equation
\begin{equation} \label{DPformulation}
  \frac{dx}{dt} = L(x) \frac{\partial E(x)}{\partial x} + \left. \frac{\partial \Psi^*(x,\xi)}{\partial \xi} \right|_{\xi=\frac{\partial S(x)}{\partial x}} \, .
\end{equation}
In Eq.~(\ref{DPformulation}), the dissipation potential $\Psi^*$ is a convex real-valued function of $\xi$ that has its minimum at $0$, where $\Psi^*(x,0)=0$  (for the origins of using dissipation potentials in irreversible thermodynamics, see \cite{Edelen72,vanKampen73,Grmela84ip,Grmela93a,Grmela93b}; see also the remarks in Sec.~2.9 of \cite{Grmela18}). The potential $\Psi^*$ can have an additional explicit dependence on $x$. The formulation of irreversible dynamics in terms of a dissipation potential is also known as generalized gradient flow (for details, see \cite{Mielkeetal14} and references therein). In the following, we hence distinguish between the GENERIC implementation of irreversible dynamics by gradient flows in Eq.~(\ref{LMformulation}) and by generalized gradient flows in Eq.~(\ref{DPformulation}); equivalently, we refer to gradient flows based on friction matrices and based on dissipation potentials. For symmetric $M(x)$, the quadratic dissipation potential $\Psi^*(x,\xi) = (1/2) \, \xi \cdot M(x) \xi$ reproduces the gradient flow appearing in Eq.~(\ref{LMformulation}).

The respective advantages of the formulations (\ref{LMformulation}) and (\ref{DPformulation}) have been discussed in the literature, in most detail in \cite{HutterSvendsen13}.
For the formulation of irreversible dynamics based on dissipation potentials, the strong formulation of energy conservation and problems associated with dimensional arguments of inhomogeneous functions have been addressed in \cite{hco236}, where, based on physical arguments, also the combination of the formulations (\ref{LMformulation}) and (\ref{DPformulation}) has been advocated and elaborated.

A preference for the formulation (\ref{LMformulation}) or (\ref{DPformulation}) has often been considered as a matter of taste. Only Eq.~(\ref{LMformulation}) can handle Casimir symmetry, whereas Eq.~(\ref{DPformulation}) is clearly more natural for chemical reactions and for collisions in the Boltzmann equation. In the present paper, we show that the choice of a formulation actually implies the type of noise by which a phenomenological equation can be enhanced. The choice should be based on the physical situation and hence be made with deliberation.

\section{Fluctuation-dissipation theorem}\label{secFDT}
The fluctuation-dissipation theorem plays a key role in describing nonequilibrium systems. As a general principle, it was first formulated by Nyquist \cite{Nyquist28} in 1928 and later derived by Callen and Welton \cite{CallenWelton51}, but important special cases had been noted in the preceding decades. Kubo and coworkers \cite{KuboetalII} proposed a scheme for classifying the various types of formulas referred to as fluctuation-dissipation relations.

Here we consider the \emph{fluctuation-dissipation theorem} as the principle that establishes a relationship between the (generalized) gradient flows characterizing the irreversible contribution to GENERIC on the one hand and the appropriate fluctuation enhancement of these thermodynamically structured equations on the other hand. The Markov processes providing the fluctuation enhancement can be characterized in various ways: by stochastic differential equations, transition rates, functional integrals, infinitesimal generators, or by nonlinear generators.

\subsection{Fluctuations associated with friction matrices}\label{secflucfricmat}
A heuristic derivation of the GENERIC (\ref{LMformulation}) based on friction matrices from more microscopic equations is available (see, for example, Chapt.~6 of \cite{hcobet}). A well-established procedure is based on the projection-operator technique for separating slow and fast dynamics \cite{Mori65,Zwanzig61,Robertson66,Grabert}, where the slow dynamics provides the phenomenological equations and the fast dynamics is idealized as stochastic noise. Adaptations of the general ideas of the projection-operator technique to GENERIC can be found in \cite{hco101} for classical systems and in \cite{hco131} for quantum systems (see also App.~D of \cite{hcobet}).

The projection-operator technique suggests the following fluctuation-enhancement procedure for the version of GENERIC given in Eq.~(\ref{LMformulation}) (see, e.g., Eq.~(1.56) and Sec.~6.3.3 of \cite{hcobet}),
\begin{multline}\label{LMformulationfluc}
  dX_t = \bigg( L(x) \frac{\partial E(x)}{\partial x} + M(x) \frac{\partial S(x)}{\partial x} \\
  + \kB \frac{\partial}{\partial x} \cdot M(x) \bigg)\bigg|_{x=X_t} dt + B(X_t) \, dW_t \, ,
\end{multline}
where $B$ is a (not necessarily square) matrix satisfying
\begin{equation}\label{BCholesky}
   B(x) B(x)^T = 2 \kB M(x) \, ,
\end{equation}
and $\kB$ is Boltzmann's constant. Different choices of the matrix $B$ in the decomposition (\ref{BCholesky}) correspond to different but equivalent versions of the Wiener process $W_t$ in Eq.~(\ref{LMformulationfluc}), which is a vector-valued Gaussian stochastic process with the following first and second moments,
\begin{equation}\label{Wienermoms}
   \mathbb{E}(W_t) = 0 \, , \quad \mathbb{E}(W_t W_{t'}^T) = \min(t,t') \, \mathbbm{1} \, .
\end{equation}
The smallness of the Boltzmann constant in macroscopic units of entropy, $\kB = 1.38 \times 10^{-23} \, {\rm J/K}$, highlights the microscopic origin and smallness of the fluctuation effects characterized by Eq.~(\ref{BCholesky}). Equation (\ref{LMformulationfluc}) establishes a relationship between the GENERIC phenomenological equations based on friction matrices and the class of stochastic processes known as diffusions (the continuous solutions to stochastic differential equations) and shows that a friction matrix~$M$ contains all the essential information to reconstruct the fluctuations that arise from neglected more microscopic degrees of freedom. We occasionally refer to Eqs.~(\ref{LMformulationfluc}) and (\ref{BCholesky}) as the classical fluctuation-dissipation theorem. Note that temperature does not appear in these equations. Energy and entropy are the fundamental concepts in GENERIC, whereas nonequilibrium temperature is not even defined in a meaningful general way. However, in the GENERIC formulation of hydrodynamics, for example, the local-equilibrium temperature occurs as a factor in the friction matrix $M$.

The occurrence of the divergence of $M$ in Eq.~(\ref{LMformulationfluc}) is a consequence of using the It\^o interpretation of stochastic differential equations with multiplicative noise. This term may be regarded as a warning that the phenomenological equation should not be considered naively as an averaged version of the stochastic differential equation; rather, their solutions correspond to the \emph{most probable} paths. Without the correction term, the infinitesimal generator characterizing the solution of the stochastic differential equation (\ref{LMformulationfluc}) would be
\begin{equation}\label{diffusiongenerator}
   {\cal Q} = \left( L \frac{\partial E}{\partial x} + M \frac{\partial S}{\partial x} \right) \cdot \frac{\partial}{\partial x} + \kB M : \frac{\partial}{\partial x} \frac{\partial}{\partial x} \, .
\end{equation}
With the correction term, the infinitesimal generator is changed into
\begin{equation}\label{diffusiongeneratorsym}
   {\cal Q} = \left( L \frac{\partial E}{\partial x} + M \frac{\partial S}{\partial x} \right) \cdot \frac{\partial}{\partial x} + \frac{\partial}{\partial x} \cdot \kB M \cdot \frac{\partial}{\partial x} \, .
\end{equation}

We can rewrite Eq.~(\ref{LMformulationfluc}) in the more appealing form
\begin{align}
  	dX_t &= \bigg( L(x) \frac{\partial E(x)}{\partial x} + M(x) \frac{\partial S(x)}{\partial x} \bigg)\bigg|_{x=X_t} dt \nonumber\\
  	&+ B(X_t) \diamond dW_t \, , \label{LMformulationflucx}
\end{align}
where the symbol $\diamond$ indicates the kinetic or Klimontovich interpretation of this stochastic differential equation \cite{Klimontovich90,hco113}. As a general policy, we keep the general form of equations (here stochastic differential equations, later also functional integrals) as simple as possible, but emphasize that these formal equations require interpretation rules, most pragmatically expressed through time-discretization schemes.

By passing from discrete systems to the limit of fields, Eq.~(\ref{LMformulationfluc}) can be used to add fluctuations to hydrodynamics, resulting in the famous theory of fluctuating hydrodynamics \cite{LandauLifshitz9}. Note that this limiting procedure comes with serious difficulties, which are addressed in the theory of stochastic partial differential equations \cite{DaPratoZabczyk,KipnisLandim,Espanol98}.

Equation~(\ref{LMformulationfluc}) or (\ref{LMformulationflucx}) provides the fluctuation enhancement in terms of a stochastic differential equation. The same fluctuation enhancement can be described unambiguously by the infinitesimal generator (\ref{diffusiongeneratorsym}). Alternatively, one can provide the probability distribution of the corresponding stochastic process in the space of trajectories. In that approach, which has been pioneered by Onsager and Machlup \cite{OnsagerMachlup53}, averages are obtained as functional integrals, also known as path integrals. Considering the solution of the stochastic differential equation (\ref{LMformulationfluc}) on the interval $[0,T]$, the probability density in the space of trajectories is given by
\begin{multline}\label{OnsagerMachlupP}
   \mathbb{P}(X_t \approx x_t) \propto \\
   p(x_0) \, \exp\left\{ - \frac{1}{2\kB} \int_0^T \mathcal{F}\!\left( x_t, \dot{x}_t - L(x_t) \frac{\partial E(x_t)}{\partial x_t} \right) dt \right\} ,
\end{multline}
where
\begin{multline}\label{OnsagerMachlupF}
   {\cal F}(x,v) = \frac{1}{2} \left( v - M(x) \frac{\partial S(x)}{\partial x} \right) \cdot \\
   \cdot M(x)^{-1} \left( v - M(x) \frac{\partial S(x)}{\partial x} \right) .
\end{multline}
The meaning of the inverse $M^{-1}$ of a degenerate matrix $M$ in the Gaussian probability density (\ref{OnsagerMachlupP}), (\ref{OnsagerMachlupF}) requires some explanation. If the second moments of $v$ are degenerate, this means that the fluctuations of $v$ are constrained. A regularization of $M$ can be obtained by introducing small artificial Gaussian fluctuations in the constrained directions, so that the inverse of the regularized $M$ exists. After calculating the desired averages, one should let these small artificial fluctuations go to zero.

The meaningful definition of formal functional integrals is a subtle matter. In particular, the treatment of $\dot{x}_t$ in Eq.~(\ref{OnsagerMachlupP}) should put initial and final times on an equal footing. Interpretation rules for functional integrals have been developed in \cite{CugliandoloLecomte17,ItamiSasa17}. The need to introduce interpretation rules for functional integrals should not keep us from taking advantage of this powerful tool, as is well-known from a similar situation in the theory of stochastic differential equations.

Note that the function defined in Eq.~(\ref{OnsagerMachlupF}) has the property ${\cal F}(x,v) \geq 0$ and that its minimum at zero is reached if $v=M \partial S/\partial x$ is the irreversible contribution to dynamics in the GENERIC (\ref{LMformulation}). This means that the time integral in the probability density (\ref{OnsagerMachlupP}) vanishes if $x_t$ satisfies the phenomenological equation, so that this trajectory has the maximum probability. Stochastic deviations from the deterministic solution are suppressed. In view of the small value of $\kB$, only microscopic fluctuations have nonnegligible probabilities, whereas macroscopically large deviations are exponentially suppressed.

\subsection{Fluctuations associated with dissipation potentials}\label{secflucdispot}
The microscopic justification of dissipation potentials is much less developed than for friction matrices. A mathematical derivation from Hamiltonian dynamics based on an optimization principle has been offered by Turkington \cite{Turkington13}. The fluctuations in reactive mixtures have been discussed in \cite{Bhattacharjeeetal15,Kimetal17}. The elegance and generality of the GENERIC (\ref{DPformulation}) based on dissipation potentials suggests that all these pieces of a puzzle can be put together to obtain a powerful theory of fluctuations. General results for purely dissipative systems, in the language of large-deviation theory, have been revealed for the class of Markov processes, including jump processes, in \cite{Mielkeetal14,Mielkeetal16,Kraaijetal19}, from which our work originates. The variational principle ${\cal F}(x,v) \geq 0$ (where ${\cal F}(x,v) = 0$ corresponds to the deterministic phenomenological equation) is the key to a general description of fluctuations in terms of functional integrals.

As a convex function of $\xi$, the dissipation potential $\Psi^*(x,\xi)$ gives naturally rise to another convex function $\Psi(x,v)$, where $\Psi(x,v)$ and $\Psi^*(x,\xi)$ are related by Legendre transformation. The more general Legendre-Fenchel transform \cite{Rockafellar} is defined by
\begin{equation}\label{LegendreFenchel}
   \Psi(x,v) = \sup_\xi \, [\xi \cdot v - \Psi^*(x,\xi)] \, .
\end{equation}
If the dissipation potential $\Psi^*(x,\xi)$ is sufficiently smooth, the more familiar Legendre transform can be rewritten as
\begin{equation}\label{Legendre}
   \Psi(x,v) = \xi \cdot v - \Psi^*(x,\xi) \, ,
\end{equation}
where the variables $v$ and $\xi$ are related by
\begin{equation}\label{Legendrexiv1}
   v = \frac{\partial \Psi^*(x,\xi)}{\partial \xi} \, ,
\end{equation}
or, equivalently,
\begin{equation}\label{Legendrexiv2}
   \xi = \frac{\partial \Psi(x,v)}{\partial v} \, .
\end{equation}
The physical meaning of the conjugate variables $v$ and $\xi$ is provided by Eq.~(\ref{DPformulation}),
\begin{equation}\label{Legendrexivphys}
   v = \frac{dx}{dt} - L(x) \frac{\partial E(x)}{\partial x} \, , \qquad \xi = \frac{\partial S(x)}{\partial x} \, ,
\end{equation}
which are the irreversible contribution to dynamics and the entropy gradient, respectively. This interpretation of the variables is based on the phenomenological equation; in the presence of fluctuations, deviations from the values (\ref{Legendrexivphys}) occur.

The definition (\ref{LegendreFenchel}) implies that, for any $x$, $\xi$, and $v$, we have the Young-Fenchel inequality
\begin{equation}\label{YoungFenchelineq}
   \Psi(x,v) + \Psi^*(x,\xi) - \xi \cdot v \geq 0 \, .
\end{equation}
If $\xi$ and $v$ are related according to Eq.~(\ref{Legendrexiv2}), in particular for the physical variables (\ref{Legendrexivphys}), equality is reached. We thus obtain the second law of thermodynamics,
\begin{equation}\label{Sgrowthdispot}
   \frac{dS}{dt} = \frac{\partial S}{\partial x} \cdot \frac{dx}{dt} = \frac{\partial S}{\partial x} \cdot v = \Psi(x,v) + \Psi^*\left(x,\frac{\partial S}{\partial x}\right) \geq 0 \, .
\end{equation}
In the second step, we have used the degeneracy of the Poisson matrix $L$. Finally, both $\Psi(x,v)$ and $\Psi^*(x,\xi)$ are nonnegative.

The following generalization of the function ${\cal F}(x,v)$ defined in Eq.~(\ref{OnsagerMachlupF}) turns out to be useful:
\begin{equation}\label{generalFdef}
   {\cal F}(x,v) = \Psi(x,v) + \Psi^*(x,\xi) - \xi \cdot v \Big|_{\xi=\frac{\partial S(x)}{\partial x}} \, .
\end{equation}
This function inherits the convexity in $v$ from $\Psi(x,v)$. The inequality (\ref{YoungFenchelineq}) implies ${\cal F}(x,v) \geq 0$. The minimum of ${\cal F}(x,v)$ is reached for the variables (\ref{Legendrexivphys}), that is, for the solutions of the GENERIC (\ref{DPformulation}) based on dissipation potentials. This means that the solutions of the macroscopic, GENERIC phenomenological equations are the most probable paths of the stochastic process. We have thus arrived at a variational principle associated with Eq.~(\ref{DPformulation}). This variational principle is somewhat unusual because we know in advance that the minimum is reached at zero. In the special case of quadratic dissipation potential, the trivial nature of the variational principle associated with minimizing ${\cal F}(x,v)$ in Eq.~(\ref{OnsagerMachlupF}) becomes evident.

Nevertheless, the function ${\cal F}(x,v)$ is very useful for specifying the fluctuation-enhanced version of the GENERIC (\ref{DPformulation}) based on generalized gradient flows. The probability density in the space of trajectories is still given by Eq.~(\ref{OnsagerMachlupP}), but now ${\cal F}(x,v)$  is obtained from the dissipation potential according to Eq.~(\ref{generalFdef}). With this relation between path integrals and dissipation potentials, we have arrived at a first formulation of the generalized fluctuation-dissipation theorem, which we consider as a cornerstone of nonequilibrium statistical mechanics.
\begin{FDT}[first formulation]
 A fluctuation enhancement of a GENERIC system with building blocks $E$, $S$, $L$, and $\Psi^*$ is completely characterized by the probability density (\ref{OnsagerMachlupP}) given in terms of the function $\mathcal{F}$ defined in Eq.~(\ref{generalFdef}).
\end{FDT}
While, according to Einstein's fluctuation theory for enhancing equilibrium thermodynamics, an entropy function describes the fluctuations around most probable states in \emph{static} situations (systems that do not evolve macroscopically in time), a dissipation potential characterizes the fluctuations around the most probable \emph{dynamics} of a system. Such fluctuations are characterized by transport coefficients and/or rate parameters which should not occur in entropies or free energies \cite{hco191}. Note that, as reflected by the existence of the local function $\mathcal{F}(x, v)$, the formulation refers to the class of Markov processes, which are \emph{local in time}. Any \emph{nonlocal-in-space} feature is, in principle, allowed.

\subsection{Infinitesimal generators}
The integral in the exponent of Eq.~(\ref{OnsagerMachlupP}) encodes the sequential transition probabilities of a Markov process (see \cite{CugliandoloLecomte17,ItamiSasa17} or Sec.~11.1 of \cite{Honerkamp}). Therefore, the infinitesimal generator ${\cal Q}$ of this Markov process, which is defined as the operator
\begin{equation}\label{infgeneratordef}
   {\cal Q} f(x) = \lim_{\tau \rightarrow 0} \frac{1}{\tau} \mathbb{E}\Big[ f(X_\tau) - f(x) \Big| X_0=x \Big] \, ,
\end{equation}
for an appropriate class of functions $f$, can be expressed as
\begin{multline}\label{infgeneratorp}
   {\cal Q} f(x) = \\
   \lim_{\tau \rightarrow 0} \frac{1}{\tau} \int \left[ f \left( x + L \frac{\partial E}{\partial x} \tau + v \tau \right) - f(x) \right]   p^\tau_x(v) \, dv \, , \\
\end{multline}
with the short-time transition probabilities
\begin{equation}\label{transitionpdef}
   p^\tau_x(v) = \exp\left\{ - \frac{{\cal F}(x,v)}{2\kB} \tau \right\} \left[ \int \exp\left\{ - \frac{{\cal F}(x,v)}{2\kB} \tau \right\}  dv \right]^{-1} .
\end{equation}
Equation (\ref{infgeneratorp}) corresponds to a forward Euler, or nonanticipating, definition of the functional integral in Eq.~(\ref{OnsagerMachlupP}) (see \cite{CugliandoloLecomte17,ItamiSasa17} or Sec.~11.1 of \cite{Honerkamp}). The simplest way to achieve a symmetric treatment of initial and final times in a time step $\tau$ is by using the following replacement in the numerator and denominator of Eq.~(\ref{transitionpdef}),
\begin{equation}\label{transitionpdefsym}
   {\cal F}(x,v) \rightarrow {\cal F}(x + v\tau/2,v) \, ,
\end{equation}
where typically a first-order Taylor expansion of the symmetrizing correction is sufficient.

Equations (\ref{infgeneratorp}) and (\ref{transitionpdef}) characterize a Markov process and may be regarded as an alternative formulation of the general fluctuation-dissipation theorem (\ref{OnsagerMachlupP}), (\ref{generalFdef}). The characterization of a Markov process via its infinitesimal generator is mathematically more robust than via a functional integral (contrary to stochastic and functional integrals, infinitesimal generators do not require interpretation rules). In view of Eq.~(\ref{generalFdef}), we find the following alternative representation of the short-time transition probabilities,
\begin{align}\label{transitionpdefPsi}
   p^\tau_x(v) &= \exp\left\{ \frac{\tau}{2\kB} \left( \frac{\partial S(x)}{\partial x} \cdot v - \Psi(x,v) \right) \right\} \nonumber \\
   &\times \left[ \int \exp\left\{ \frac{\tau}{2\kB} \left( \frac{\partial S(x)}{\partial x} \cdot v - \Psi(x,v) \right) \right\} dv \right]^{-1} . \nonumber \\
\end{align}

For the quadratic expression (\ref{OnsagerMachlupF}) of ${\cal F}(x,v)$, the short-time transition probabilities (\ref{transitionpdef}) are Gaussians with the first moments and covariances
\begin{equation}\label{quadraticFmoms}
   \langle v \rangle^\tau_x = M(x) \frac{\partial S(x)}{\partial x} \, , \quad\! \langle v v^T \rangle^\tau_x - \langle v \rangle^\tau_x \, \langle v^T \rangle^\tau_x = \frac{2 \kB M(x)}{\tau} \, .
\end{equation}
The fact that the fluctuating velocities $v$ are of order $\tau^{-1/2}$ expresses the non-differentiability of the continuous trajectories of diffusion processes. In the limit $\tau \rightarrow 0$, a second-order Taylor expansion of the first $f$ in Eq.~(\ref{infgeneratorp}) around $x$ is sufficient to obtain the infinitesimal generator (\ref{diffusiongenerator}). If the correction (\ref{transitionpdefsym}) is applied, the generator changes to (\ref{diffusiongeneratorsym}), which expresses the classical fluctuation-dissipation theorem.

In short, we find that a nonanticipating definition of the functional integral leads to the infinitesimal generator (\ref{diffusiongenerator}), whereas a symmetric definition leads to the generator (\ref{diffusiongeneratorsym}). As for stochastic integrals, the definition of functional integrals depends on the discretization rule. For stochastic integrals, symmetrization leads from It\^o to Stratonovich integrals whereas, for functional integrals, symmetrization leads from It\^o to Klimontovich generators.

For general ${\cal F}(x,v)$, we deal with jump processes so that $v$ is of order $\tau^{-1}$. Then, all orders of the Taylor expansion for $f$ in Eq.~(\ref{infgeneratorp}) contribute in the limit $\tau \rightarrow 0$ (truncating after one or two terms or keeping infinitely many terms are the only options; see, e.g., Sec.~4.3 of \cite{Risken}). To handle the infinite-order expansion, we consider the test functions $f(x)=e^{\alpha \cdot x}$ and, for the average occurring in the infinitesimal generator (\ref{infgeneratorp}), we then obtain
\begin{multline}\label{generatorexps}
   \left\langle f \left( x + L \frac{\partial E}{\partial x} \tau + v \tau \right) \right\rangle^\tau_x  = \\
   f(x) \, \exp\left\{ \alpha \cdot L \frac{\partial E}{\partial x} \tau \right\} \left\langle e^{\alpha \cdot v \tau} \right\rangle^\tau_x \, .
\end{multline}
We now evaluate the integrals involved in the remaining average $\left\langle e^{\alpha \cdot v \tau} \right\rangle^\tau_x$ by the method of steepest descent. This may be surprising in view of the smallness of the time step $\tau$. However, it actually is the smallness of $\kB$ that justifies the application of the method of steepest descent. More precisely, the entropy changes involved in the fluctuations during $\tau$ should still be large compared to $\kB$, which means that $\tau$ should be sufficiently large so that many degrees of freedom are involved in these fluctuations. This assumption is a key requirement for applying statistical mechanics. By using the representation (\ref{transitionpdefPsi}) of $p^\tau_x(v)$ and applying the method of steepest descent, we obtain
\begin{equation}\label{steepestdescentres}
   \left\langle e^{\alpha \cdot v \tau} \right\rangle^\tau_x \sim \frac{\exp\left\{ \frac{\tau}{2\kB} \Psi^*(x,\xi) \right\}}{\exp\left\{ \frac{\tau}{2\kB} \Psi^*\left(x,\frac{\partial S(x)}{\partial x}\right) \right\}} \, ,
\end{equation}
with
\begin{equation}\label{fluxxichoice}
   \xi = \frac{\partial S(x)}{\partial x} + 2 \kB \alpha \, .
\end{equation}
In evaluating the normalization integral in Eq.~(\ref{transitionpdefPsi}), stationarity corresponds to the physical value of $\xi$ given in Eq.~(\ref{Legendrexivphys}) for the deterministic phenomenological evolution. The stationarity condition for the integral in the numerator implies a deviation characterized by Eq.~(\ref{fluxxichoice}). Note that in applying the method of steepest descent we have omitted the ratio of determinants of the matrix of second derivatives of $\Psi(x,v)$ with respect to $v$, which needs to be evaluated for different $v$. For quadratic dissipation potential, the second derivatives are independent of $v$. More generally, the correction factor resulting from the determinants is very small compared to the large exponentials in Eq.~(\ref{steepestdescentres}) [the same is true for the correction (\ref{transitionpdefsym}) for time-reversal symmetry]. A most useful, compact result is obtained after taking logarithms in Eqs.~(\ref{generatorexps}) and (\ref{steepestdescentres}),
\begin{multline}\label{centralformula}
   \frac{2 \kB}{\tau} \ln \left\langle \exp\left\lbrace \alpha \cdot \left( L \frac{\partial E}{\partial x} \tau + v \tau \right) \right\rbrace \right\rangle^\tau_x \approx \\
   \Psi^*(x,\xi) - \Psi^*\left(x,\frac{\partial S(x)}{\partial x}\right) + \xi \cdot  L(x) \frac{\partial E(x)}{\partial x} \, ,
\end{multline}
where $\xi$ is obtained from $\alpha$ according to Eq.~(\ref{fluxxichoice}). A comparison with Eq.~(\ref{infgeneratorp}) suggests interpreting the left-hand side of Eq.~(\ref{centralformula}) as a nonlinear version of the infinitesimal generator \cite{FengKurtz}. Note, however, that the validity of Eq.~(\ref{centralformula}) is based on the smallness of $\kB/\tau$ compared to the macroscopic entropy production rate on the intermediate time scale $\tau$, and not on a formal limit $\tau \rightarrow 0$. The seemingly conflicting requirements for $\tau$ in using the method of steepest descent and in defining a nonlinear infinitesimal generator characterize $\tau$ as an intermediate time scale, large compared to microscopic and small compared to macroscopic time scales. A more formal treatment of precisely these complementary limits can be obtained by means of large-deviation theory.

The average in Eq.~(\ref{centralformula}) is basically the characteristic function (the Fourier transform of the probability measure) associated with the fluctuating increments of the trajectory over the time interval $\tau$ (if $\alpha$ is imaginary). As $\tau$ is small, the increments are small, the characteristic function is close to unity, and its logarithm is close to zero; this explains why it is meaningful to divide by $\tau$ in Eq.~(\ref{centralformula}). For diffusion processes, the increments are Gaussian and the logarithm of the characteristic function, which is the cumulant generating function, is a quadratic function in $\alpha$,
\begin{equation}\label{cumulantgen}
	\alpha \cdot \left( L \frac{\partial E}{\partial x} + M \frac{\partial S}{\partial x} \right) \tau + \alpha \cdot M \cdot \alpha \, \kB \tau \, .
\end{equation}
The first and second cumulants, which are the only nonvanishing ones, are both proportional to $\tau$. In more general situations, we obtain \eqref{centralformula}, which constitutes our second formulation of the FDT.
\begin{FDT}[second formulation]
A fluctuation enhancement of a GENERIC system with building blocks $E$, $S$, $L$, and $\Psi^*$ is completely characterized by the cumulant generating function (\ref{centralformula}), where $\xi$ is defined in Eq.~(\ref{fluxxichoice}).
\end{FDT}

Note that the right-hand side of Eq.~(\ref{centralformula}) coincides with the contact Hamiltonian given in Eq.~(47) of \cite{hco99}. This contact Hamiltonian is actually the Legendre-Fenchel transform of the Lagrangian (\ref{generalFdef}), and the corresponding contact Hamiltonian dynamics is equivalent to the evolution equations obtained from the variational principle associated with this Lagrangian \cite{MerkerKruger13,Goto15,BravettiCruzTapias17,WangWangYan17}.

\section{Recipes for nonequilibrium statistical mechanics}\label{secstatmech}
In the context of GENERIC with friction matrices or fluctuations modeled by diffusion processes, the four building blocks characterizing the structure of the phenomenological equations can be obtained by statistical mechanics as elaborated in Chapt.~6 of \cite{hcobet} or in \cite{hco173}. The generators energy~$E$ and entropy~$S$ of reversible and irreversible dynamics, respectively, as well as the Poisson matrix~$L$, require only static information that can be obtained most efficiently from Monte Carlo simulations of nonequilibrium ensembles. Dynamic simulations are required for evaluating the friction matrix~$M$. In view of the cumulant generating function (\ref{cumulantgen}), we can extract $M$ from
\begin{equation}
    \tau \left\langle v v^T \right\rangle^\tau_x - \tau \langle v \rangle^\tau_x \, \langle v^T \rangle^\tau_x = 2 \kB M(x) \, .
\end{equation}
In practice, the left-hand side is evaluated for small time increments $\tau$ and, for this reason, we may just compute
\begin{equation}\label{GreenKubo}
	\tau \left\langle v v^T \right\rangle^\tau_x = 2 \kB M(x) \, ,
\end{equation}
since the term~$\tau \langle v \rangle^\tau_x \, \langle v^T \rangle^\tau_x$ is of order $\tau^1$ whereas $\tau \left\langle v v^T \right\rangle^\tau_x$ is of order $\tau^0$. In a simulation, one creates an ensemble of microscopic initial conditions that are consistent with the thermodynamic variables $x$ and that are evolved with the microscopic dynamics. This procedure for evaluating second moments clarifies what we mean by the ``noise idealization of fast dynamics'' and by ``fluctuation-enhanced phenomenological dynamics.'' As discussed in the justification of the method of steepest descent in the paragraph before Eq.~(\ref{steepestdescentres}), $\tau$ should be small from the macroscopic perspective, but large from a microscopic perspective. More precisely, for diffusion processes, this means that $\tau$ should be the separating time scale between the fast processes idealized as noise and the slow processes described by thermodynamic evolution equations.

Dynamic simulations are performed only over the intermediate time scale $\tau$, not over macroscopic times. This is where the efficiency of thermodynamically guided simulations comes from \cite{hco173}. If the friction matrix $M$ is independent of $x$, it is convenient to average over an ensemble of initial conditions for different $x$, for example, an equilibrium ensemble. The above procedure for diffusion processes is closely related to the more common approach based on Green-Kubo formulas \cite{KuboetalII,EvansMorriss}.

Statistical mechanics for GENERIC with dissipation potential or fluctuations modeled by general Markov processes is based on Eq.~(\ref{centralformula}). For a given $x$, one needs to choose $\alpha$ and $\tau$. The choice of $\alpha$ is determined by the range of $\xi$ values of interest according to Eq.~(\ref{fluxxichoice}). The intermediate time scale $\tau$ should be such that a small but nonnegligible fraction of the simulated trajectories contain a jump during the period $\tau$.

The left-hand side of Eq.~(\ref{centralformula}) can be evaluated in exactly the same way as the second moment in the Green-Kubo formula (\ref{GreenKubo}). A fit to the right-hand side provides the noise idealization for the neglected degrees of freedom and the resulting dissipation potential according to the generalized fluctuation-dissipation theorem. If the result is plotted for fixed $x$ as a function of $\xi$, according to Eq.~(\ref{centralformula}), we obtain the dissipation potential $\Psi^*(x,\xi)$, except for constant and linear contributions. For $\xi=\partial S/\partial x$, the right-hand side is zero; for $\xi=0$, the right-hand side is $-\Psi^*(x,\partial S/\partial x)$. The static building blocks $E$, $S$, and $L$ can be extracted from molecular simulations exactly as described before for GENERIC based on friction matrices.

\section{Summary and outlook}
The main result of the present paper is a very general fluctuation-dissipation theorem that relates the dissipative contribution in a thermodynamically consistent evolution equation to a fluctuation enhancement of that equation in terms of a Markov process. In its most general form, dissipative dynamics is defined in terms of a dissipation potential $\Psi^*$ in Eq.~(\ref{DPformulation}). The Markov process providing the fluctuation enhancement of the phenomenological Eq.~(\ref{DPformulation}) can be characterized in several equivalent ways (under suitable assumptions):
\begin{itemize}
  \item By the probability distribution (\ref{OnsagerMachlupP}) in the space of trajectories, where the general Lagrangian ${\cal F}$ is given in Eq.~(\ref{generalFdef});
  \item By the infinitesimal generator (\ref{infgeneratorp}) associated with the transition probabilities $p^\tau_x$ given in Eq.~(\ref{transitionpdef}) or Eq.~(\ref{transitionpdefPsi}), where the correction (\ref{transitionpdefsym}) should be applied;
  \item By a nonlinear infinitesimal generator obtained as a limiting case of Eq.~(\ref{centralformula}) for small time steps $\tau$ (compared to macroscopic time scales).
\end{itemize}
Our generalized fluctuation-dissipation theorem has the status of a conjecture for which there is no rigorous mathematical proof. For the quadratic dissipation potentials associated with diffusion processes, projection-operator techniques can be used for a heuristic derivation. For the more general dissipation potentials associated with jump processes, there are only few deeper justifications \cite{Bhattacharjeeetal15,Kimetal17,Mielkeetal14}.

The fact that fluctuations and dissipation are two sides of the same coin can be exploited in different ways which, when properly combined, lead to a consistent multiscale description of nonequilibrium systems. On the one hand, phenomenological evolution equations rooted in nonequilibrium thermodynamics can be enhanced by introducing fluctuations representing the essential effects of neglected microscopic degrees of freedom. Important examples include the hydrodynamic equations for the discussion of dynamic light scattering and microbead rheology. On the other hand, by analyzing the fluctuations of microscopic simulations, one obtains recipes for extracting the dynamic material information expressed through dissipation potentials. The general recipe of nonequilibrium statistical mechanics is contained in Eq.~(\ref{centralformula}). This equation unifies and generalizes some famous results: the Green-Kubo formulas for transport coefficients (associated with gradient flows, friction matrices, and continuous noise) and the Kramers formula for chemical reactions rates (associated with generalized gradient flows, dissipation potentials, and jump processes).

In the subsequent paper, we illustrate how the general recipes of nonequilibrium statistical mechanics can be applied to obtain the dissipation potential for purely dissipative systems and, in particular, for a simple chemical reaction. Applications of greater practical relevance will require the development of sophisticated importance-sampling techniques.

\bibliographystyle{apsrev4-2}

\end{document}